\documentstyle[12pt,psfig,epsf]{article}
\begin{document}
\title{Isgur--Wise function in a relativistic model of
constituent quarks} \author{ A.~F.~Krutov\thanks{E-mail:
krutov@ssu.samara.ru},
O.~I.~Shro\thanks{E--mail:oshro@ssu.samara.ru}\\
{\small {\em Samara State University, 443011, Samara, Russia
}}\\
V.~E.~Troitsky\thanks{E-mail:
troitsky@theory2.npi.msu.su}\\
{\small {\em Nuclear Physics
Institute, Moscow State University, 119899, Moscow, Russia }} }
\date{November 6, 2000}
\maketitle

\begin{abstract}
The integral representation for Isgur -- Wise function (IWF) is obtained in
the framework of instant--form relativistic Hamiltonian
dynamics for mesons with one heavy quark. The upper and lower limits are
calculated for the slope parameter
of IWF $\rho^2$ by the model independent way. IWF is calculated for
different wave functions
of quarks in the meson.
The difference between the limits of $\rho^2$
equals 1/3. The constraint on the slope parameter is in a good agreement with
experiments.
The weak dependence of IWF on the choice of wave functions
is found.
\end{abstract}

{PACS numbers:  12.39.Ki, 12.60.Rc, 13.20.He}

\noindent

Key words: constituent quark model, semileptonic decays, Isgur--Wise function
\bigskip

The quark composite systems with one heavy quark have been the focus
of attention for the last several years.
In the infinite heavy quark mass limit the additional spin-flavour symmetries
appear and
simplify the calculation of matrix elements of electroweak transitions
\cite{IsW8990, Neu9293}.
Therefore all mesons form factors of the
electroweak transitions are determined by a single universal
Isgur -- Wise function (IWF)
$\xi_{IW}(w)$ \cite{IsW8990} ( $w = (v\,\cdot v')$,
$v'_{\mu}$ and $v_{\mu}$ are 4--velocities of initial and final meson).

The calculation of this function is nonperturbative in principle.
Several approaches exist to this problem,
for example, the calculations on lattices \cite{Bow95},
QCD sum rules~\cite{Neu9293,Bjo90,Vol92,BaB93,BlS93,Nar94,RafTa94},
quasipotential approach~\cite{FaG95} \cite{EbF00}, and
different forms of relativistic Hamiltonian dynamics (RHD):
light-front dynamics ~\cite{ChC96, Sim96}, point \cite{Kei97} and
instant forms \cite{YaoOl9596,MoYaOl9697}.

The slope parameter
$\rho^2 = - \xi'_{IW}(w=1)$ is of special interest, while studying the
properties of IWF.
This interest, in particular, is explained by recent measurements of this
value \cite{CLEO}, \cite{LEP}, \cite{Bat00}.
By these experiments, one can
discriminate different models
or, additionally, constraint parameters of the models.
The results of calculation of the slope parameter in different approaches
differ greatly
~\cite{Neu9293,Bow95,Bjo90,BaB93,FaG95,ChC96,Sim96,YaoOl9596,MoYaOl9697}.

In this work, IWF is calculated for mesons which contain one heavy quark.
For slope parameter $\rho^2$ the strong restriction is obtained.
This result depends weakly on interaction model for quarks.
The interval of restriction on the slope parameter equals 1/3.

The calculations of IWF are performed for different
phenomenological wave functions of quarks in meson.
It is shown that IWF depends weakly enough on the choice of the
model wave functions (about 10\% at $w\>\le$ 2).

For calculation in our work we use the instant form of RHD in
the version developed by the authors~\cite{BaK95,BaK2000}( RHD
is discussed in greater detail in Ref. ~\cite{KeP91} and
references therein).  Our approach is closely related to that in
Refs.  \cite{{YaoOl9596},{MoYaOl9697}}.  However, our approach
contains some original features, e.g., the method of
construction of current transition matrix elements
\cite{BaK2000}.

In our work, IWF is calculated from the semileptonic decay of
pseudoscalar meson  form factors~\cite{BaK2000}.
The hadron part of invariant amplitude of
the semileptonic decay of pseudoscalar meson can be expressed through two
form factors $F_{\pm}$:
\begin{equation}
\langle  \vec p_c \vert J^{\mu}\vert  \vec p\,\,'_{c'}\rangle =
P_-^{\mu}\,\, F_{-}(t) + P_+^{\mu}\,\, F_{+}(t)\>,
~\label{Jmu}
\end{equation}
where
$\vec p_c \, ,\,\vec p\,\,'_{c'}$ are 3-momenta of mesons,
$P_-^{\mu} =(p_c - p'_{c'})^{\mu} \>,\>
P_+^{\mu} =(p_c + p'_{c'})^{\mu}\>,\>
t =(p_c - p'_{c'})^2 \>.$

In our approach, the form factors $F_{-}(t)\>,\>F_{+}(t)$ have the following
integral representation in the impulse approximation \cite{BaK2000}:
\begin{equation}
 F_{\pm}(t) = \int_{m_Q+m_q}^\infty
\,d\sqrt{s} \,d\sqrt{s^{\prime}} \varphi_{c} (k(s))\,
G^{(0)}_{\pm}(s,t,s^{\prime})\, \varphi_{c'} (k(s^{\prime}))\>,
~\label{F=intimp}
\end{equation}
where
$G^{(0)}_{\pm}(s,Q^2,s^{\prime})$ are the so-called free two-partical
form factors,
$\varphi_{c'}\>,\>\varphi_{c}$ are phenomenological wave functions of initial
and final mesons in the sense of RHD. They are
normalized with relativistic density states:
\begin{equation}
\varphi(k) =\sqrt{\sqrt{s}(1 - (m_Q^2 - m_q^2)^2/s^2)}\,u(k)\,k\>,
\quad
\int k^2 \,u^2(k)\, d\,k \, = \, 1\>.
~\label{wf}
\end{equation}
Where
$k(s) =\sqrt{(s^2 - 2 s (m^2_Q + m^2_q) +(m_Q^2 - m_q^2) ^2)/4\,s}$,
$m_Q\>,\>m_q$ are masses of heavy and light quarks respectively,
$u(k)$ is the non-relativistic  wave function.

Free two-partical
form factors $G^{(0)}_\pm$ have been calculated in Ref.
\cite{BaK2000}:
$$
G^{(0)}_{+}(s,t,s^{\prime})= g(s,t,s')\,
\lbrace \lbrack -\,f_3(t) \lambda (s,t,s^{\prime}) (s - s^{\prime} - t) -
$$
$$
- 2 f_1(t)\,s'\,[\,t\,(s + s^{\prime} - t) - \eta_1\,(s - s' -t) +
\eta_2\,(s - s^{\prime} + t)]
\rbrack \cos{\alpha} +
$$
$$
+ f_6(t)\,M_2\,\xi\,(s,t,s^{\prime})\,s^{\prime}\,t\,
\sin{\alpha} \rbrace\>,
$$
\begin{equation}
G^{(0)}_{-}(s,t,s^{\prime})= g(s,t,s')\,\times
~\label{G0}
\end{equation}
$$
\times
\lbrace \lbrack
f_1(t)\,[\,(s + s^{\prime} - t)^2\,(s - s^{\prime}) +
2\,s'\eta_1\,(s + 3\,s' - t)
- 2\,s'\,\eta_2(3\,s + s^{\prime} - t)] -
$$
$$
- f_3(t)\,\lambda\,(s,t,s^{\prime}) (s + s^{\prime} - t)
\rbrack \cos{\alpha}
- f_6(t)\,M_2 \xi (s,t,s^{\prime}) s^{\prime} (s - s^{\prime})
\sin{\alpha} \rbrace ,
$$
$$
g(s,t,s') = {{\sqrt{s}(s + s^{\prime} - t)
\Theta (s,t,s^{\prime})}
\over {4 \sqrt{s^{\prime}}\left[\lambda(s,t,s^{\prime})\right]^{3/2}
\sqrt{\lambda (s^{\prime},M^2_1,M^2_2)\lambda (s,M^2_3,M^2_1)}}}\>,
$$
$$
\eta_1 = M_3^2 - M_1^2\>,\quad\eta_2 = M^{2}_{2} - M^{2}_{1}\>,\quad
\eta_3 = M_2^2 - M_3^2\>,
$$
$$
\Theta (s,t,s^{\prime}) =
\theta (s - s_1) - \theta (s - s_2)\>,\>
\lambda (a,b,c) = a^2 + b^2 + c^2 -2 ( ab + bc + ac)\>,
$$
$$
\xi (s,t,s^{\prime}) =
\left\{ - M_1^2\lambda(s,t,s^{\prime}) - s s^{\prime} t -
s \eta_2(s - s^{\prime} - t) - s \eta_2^2 - s'\eta_1^2 + \right.
$$
$$
\left.+ s'\eta_1(s - s' + t) + \eta_1\,\eta_2(s + s' - t)\right\}^{1/2}.
$$
$\theta$ is Heaviside step function,
$s_1$ and $s_2$ are variables that determine the kinematically admissible
range of $s\>$ and $\>s'$ which are given by
$$
s_{1,2} = s^{\prime} + t -{1\over 2 M_2^2}(s^{\prime} + \eta_2)(t + \eta_3)
\mp {1\over 2 M_2^2} \sqrt{\lambda(s^{\prime},M_1^2,M_2^2)
\lambda(t,M_3^2,M_2^2)},
$$
$ \alpha = \omega_1 + \omega_2 + \omega_3,$
$\omega_i$ are parameters of Wigner rotation:
$$
\omega_1 = \arctan{{\xi (s,t,s^{\prime})}
\over {M_1 (s + s^{\prime}-t + 2 \sqrt{s s^{\prime}}) +
\sqrt{s s^{\prime}}
(\sqrt{s} + \sqrt{s^{\prime}}) - \eta_2\sqrt{s} - \eta_1\sqrt{s'}}}\>,
$$
$$
\omega_2 = \arctan{{\xi (s,t,s^{\prime})}
\over {M_3 (s + s^{\prime}-t + 2 \sqrt{s s^{\prime}}) + \sqrt{s}\,
(s - t + \sqrt{s s^{\prime}}) + \eta_2\sqrt{s} + \eta_1\sqrt{s'}}}\>,
$$
$$
\omega_3 =  \arctan{{\xi (s,t,s^{\prime})}
\over {M_3 s^{\prime} + M_2 (s - t) + \sqrt{s^{\prime}}
\lbrack (M_3 + M_2)^2 - t \rbrack
 + \eta_2\,(M_2 + M_3)}}\>.
$$
$M_1$ is the light quark mass,  $M_{2,3}$ are the heavy quark masses.
Quark form factors in the approximation
of pointlike quarks have a form \cite{BaK2000}:
$$
f_1(t) = (M_2 + M_3)\,f(t)\>,\quad
f_3(t) = -(M_2 - M_3)\,f(t)\>,
$$
$$
f_6(t) = -\frac{4\,M_2}{M_3^2}\,f(t),\quad
f(t) = \left[(M_3 + M_2)^2 - t \right]^{-1/2}\>\>.
$$

The matrix element of
the semileptonic decay of pseudoscalar meson can be determined in terms
of form factors $h_{\pm}(w)$ (see, for example, \cite{ChC96}):
\begin{equation}
\langle \vec p_c\vert J^{\mu}\vert \vec p\,\,'_{c'}\rangle =
\sqrt{M_c M_{c'}}
\left[h_{+}(w) (v + v')^{\mu} + h_{-}(w) (v-v')^{\mu} \right] \>.
~\label{Hmu}
\end{equation}
Here $M_{c'}, M_c$ are masses of mesons
in initial and final states.

The form factors in Eq. (\ref{Hmu}) are connected with the form factors
in Eqs. (\ref{Jmu}),
(\ref{F=intimp}) by the following formula:
\begin{equation}
h_{\pm}(w) = {{1}\over 2\sqrt{M_c M'_{c'}}}
\left[ F_{+}(t)\left( M_c \pm M'_{c'}\right)+ F_{-}(t)\left(M_c \mp M'_{c'}
\right) \right]\>.
~\label{hF}
\end{equation}
Scalar product of 4--velocities is determined through the square
of momentum transfer:\\ $w= (M^2_c + M^2_{c'} - t)/(2 M_c
M_{c'})\>.$ The IWF can be obtained from expressions
(\ref{F=intimp}), (\ref{hF}) by finding the limit
$M_2\>\sim\>M_3\>\sim\>m_Q\>\to \infty\>,\>M_1 = m_q$:
\begin{equation}
\lim_{m_Q\to\infty}h_{+}(w)= \xi_{IW}(w)\>,
~~~~~\lim_{m_Q\to\infty}h_{-}(w)= 0\>.
~\label{LinH}
\end{equation}
In our calculations, we have assumed that the wave functions in Eq.
(\ref{F=intimp}) become independent on the flavour of heavy quark and
coincide in initial and final states in limit (\ref{LinH}).

The result of calculation of IWF can be presented in following form:
\begin{equation}
\xi_{IW}(w)=\frac{1}{\sqrt{2(w+1)}}\,\int_{0}^{\infty}
\int_{-1}^{1} {k^2 \,d k \,d z}\,{\sqrt[4]{
\frac{k'^2 + m^2_q}{k^2 + m^2_q}}} \cos\,\omega
\, u (k) \, u (k')\>,
~\label{ourXi}
\end{equation}
$$k'=\left [ (\sqrt{w^2 - 1} k z + w\sqrt{k^2 + m^2_q})^2 - m^2_q
\right ]^{1/2}\>,
$$
$$
\omega = \arctan\frac{k\,\sqrt{(w^2 - 1)(1-z^2)}}
{ (w+1)m_q + \sqrt{k^2 + m^2_q} + \sqrt{k'^2 + m^2_q}}\>,
$$
The condition $\xi_{IW}(1)=1$ is a
consequence of the normalization (\ref{wf}).

In analogy with the works
\cite{Kei97,YaoOl9596,MoYaOl9697}, the slope parameter of IWF at $w=$ 1 can be
represented as the sum of three terms:
\begin{equation}
\rho^2 = \rho_{space}^2 +  \rho_{WR}^2 + \rho_{quark}^2 \>,
~\label{ourrho}
\end{equation}
\begin{equation}
\rho_{space}^2 =\frac{1}{3} \int_{0}^{\infty}\,k^2\left(k^2 + m_q^2\right)
\left(\frac{d\, u}{d\,k}\right)^2 \, d\,k \,
+\frac{1}{6} \int_{0}^{\infty} \frac{k^4 \, u^2 (k)}{k^2 + m_q^2} \, d\,k
- \frac{1}{2}\>\>,
~\label{rhos}
\end{equation}
$$
\rho_{WR}^2 =\frac{1}{6} \int_{0}^{\infty}
\frac{k^4 \, u^2 (k)}{(\sqrt{k^2 + m_q^2} + m_q)^2} \, d\,k\, \>,
~~~~~~~\rho_{quark}^2 =\frac{1}{4}\, .
$$
The terms in Eq. (\ref{ourrho}) describe the contributions of  form of
wave function
(with the consideration of relativistic normalization
(\ref{wf})), Wigner rotation, and quark current, respectively.
The contribution $\rho_{WR}^2$ and the second term in
$\rho_{space}^2$ have relativistic nature.  Let us remark that
our value of $\rho_{quark}^2$ coincides with that from
\cite{YaoOl9596}.

The integrals of the wave function can be transformed
into space coordinate representation:
$$
\int_{0}^{\infty} k^2 \left(\frac{d\, u}{d\,k}\right)^2 \, d\,k \,
= \int_{0}^{\infty} r^4 \, \psi_{0}^2 (r)\, d\,r = \left< r^2 \right>\> ,
$$
\begin{equation}
\int_{0}^{\infty} k^4 \left(\frac{d\,u}{d\,k}\right)^2 \, d\,k \, =
\int_{0}^{\infty} r^4 \left(\frac{d\, \psi_{0}}{d\,r}\right)^2 \, d\,r \, ,
~~~\int_{0}^{\infty} r^2 \, \psi_{0}^2 (r) \, d\,r = 1 .~\label{r2}
\end{equation}
Here $\left< r^2\right>$ is the mean square radius in the ground
state with wave function $\psi_{0}(r)$.

Notice that all integrals in Eqs. (\ref{ourrho}), (\ref{rhos}) are
positive.
From equations (\ref{ourrho}), (\ref{rhos}), (\ref{r2}) the restriction on
possible values of $\rho^2$ can be established.
To do so, let us consider the relativistic terms in Eqs. (\ref{ourrho}),
(\ref{rhos}) ($\rho_{WR}^2$ and the second term in $\rho_{space}^2$).

In the non-relativistic limit these positive terms
disappear in Eq. (\ref{rhos}) and we obtain the lower bound for the slope
parameter:
\begin{equation}
\rho^2 \ge \frac{1}{3} \int_{0}^{\infty} r^4
\left(\frac{d\, \psi_{0}}{d\,r}\right)^2 \, d\,r \,
+\frac{1}{3} m^2_q \left< r^2\right>
- \frac{1}{4}\>,
~\label{rhomin}
\end{equation}
In the ultrarelativistic limit these relativistic terms in Eq. (\ref{ourrho})
take the maximum
values and we obtain the upper bound for $\rho^2$:
\begin{equation}
\rho^2 \le \frac{1}{3} \int_{0}^{\infty} r^4
\left(\frac{d\, \psi_{0}}{d\,r}\right)^2 \, d\,r \, +
\frac{1}{3} m^2_q \left< r^2\right> \,
+ \frac{1}{12}\>.
~\label{rhomax}
\end{equation}
Let us remark that the difference between these maximum and minimum of the
slope parameter equals 1/3.

Now we shall use the results of works~\cite{LomMa00}.
In these works a recurrence formula was derived:
\begin{equation}
\left< r^{2 l} \right>=\frac{1}{2\,\mu} l(2 l + 1)
\frac{\left< r^{2(l-1)}\right>\>}{E_{1}(l)-E_{0}}\>.
~\label{r2n}
\end{equation}
Here $l$ is the orbital angular momentum, $\left< r^{2 l} \right>$
multipole moments in ground state of two-body system,
$E_{0}$ ground state energy, $E_{1}(l)$
energy of the lowest level of multipolarity $l\>$, $\>\mu$
reduced mass, in the limit (\ref{LinH}), $\>\mu = m_q$.
Recurrent formula (\ref{r2n})
is satisfied within $1\%$  accuracy as equality for confining potential.

For estimation of mean square radius in (\ref{rhomin}) and
(\ref{rhomax}) by means of Eq. (\ref{r2n}), we have used two measurements of
energy levels
of $D$ - meson ($1\,S_{0}$ and $1\,P_{1}$)~\cite{Gro00}.

For further estimation, let us transform integrals
in expressions (\ref{rhomin}) and (\ref{rhomax}):
\begin{equation}
\frac{1}{3} \int_{0}^{\infty} r^4
\left(\frac{d\, \psi_{0}}{d\,r}\right)^2 \, d\,r =
1 + \frac{1}{3}\int_{0}^{\infty}\psi_{0}(r)\left[(\vec r\,\hat{\vec p})^2 -
i\,(\vec r\,\hat{\vec p})\right]\psi_{0}(r)\,r^2\,d\,r\>.
~\label{Srint}
\end{equation}
Here $\hat{\vec p}$ is the operator of canonical 3-momentum of system.

For model independent estimation of integral in right part of Eq.
(\ref{Srint}),
we have used quasiclassical approximation.
In ground state, this approximation  gives $(\vec r\,{\vec p})$ = 0 because
the mean value of $r(t)$ sweeps a circle in the
quasiclassical approximation.
Therefore, the left hand side of Eq. (\ref{Srint}) is
approximately equal to unit.

For estimation of quantum corrections of this result, it is possible
to calculate exactly the integral in the left part of Eq. (\ref{Srint})
in some models. Let us perform
the calculation for the model with linear
confinement and Coulomb behavior at small distances
\cite{Tez91}:
\begin{equation}
u(r) = N_T \,e^{-\alpha r^{3/2}
- \beta r}\>, \> \alpha = \frac{2}{3}\sqrt{2\,\mu\,a}\>,\> \beta
= \mu\,b\>.
\label{Tez91-wf}
\end{equation}
In Eq.(\ref{Tez91-wf}), $a\>$ and $b$ are the parameters of linear
and Coulomb parts of potential, respectively. We use the value
of $b = 4/3\alpha_s$.  In doing so, we have used the value
$\alpha_s = 0.52$ at a scale of heavy meson masses.  The
parameter of linear part of potential $a= 0.0816\,$ GeV$^2$ was
obtained by fitting of the mean square radius in ground state of
$D$--meson which was calculated from Eq. (\ref{r2n}).  In this
way the quantum correction $\simeq$0.07 was obtained.

As follows from the general conditions of application of the
quasiclassical approximation, the less steep potential
($\sim r^\alpha\>,
\>\alpha< 1$ or $\sim\ln\,r$) must result in smaller quantum corrections.
Hence, for such class of models, the following restriction is valid to within
$\simeq$0.07 accuracy:
\begin{equation}
\frac{3}{4}\leq \rho^2 - \frac{1}{3}\, m^2_q \, \left< r^2\right>
\leq \frac{13}{12}\>.
~\label{rhoint}
\end{equation}
The light quark mass $m_q$ is now
the only free parameter in our estimations .
In our relativistic calculation in Refs. \cite{BaK95}, \cite{BaK2000}
we used the value
$m_q = 0{.}25$ GeV. This calculation are in reasonably good agreement with
experiments on electroweak properties of light mesons. In this case,
Eq. (\ref{r2n}) gives $\left< r^{2} \right>$ = 0{.}422 fm$^2$.
Hence, inequalities (\ref{rhoint}) are reduced to the following limitation:
\begin{equation}
0{.}98
\leq  \rho^2 \leq 1{.}31
~\label{rhovol}
\end{equation}
If the mass $m_q$ is varied within the limits in which the
modern relativistic calculations are performed in different
works ($0.20\>\le\>m_q\>\le\>0.33$ GeV)  then
the region of possible $\rho^2$ increases:
\begin{equation}
0{.}93 \leq  \rho^2
\leq 1{.}38
~\label{soft}
\end{equation}

The derived
restrictions are examined in our work for the following
different model wave functions:  the harmonic-oscillator
potential model (see, for example, \cite{ChC88})
\begin{equation}
u(k)= N_{HO}\,
\hbox{exp}\left(-{k^2}/{2\,b^2}\right)\>,
\label{HO-wf}
\end{equation}
the power law wave function
(see, for example, \cite{Sch94})
\begin{equation}
u(k) =N_{PL}\,{(k^2/b^2 +
1)^{-n}}\>,\quad n = 2,3\>,
\label{PL-wf}
\end{equation}
and already mentioned wave function with linear confinement
and a Coulomb behavior at small distances which is described by Eq.
(\ref{Tez91-wf}).
The wave function parameters were determined by fitting of mean
square radius of the $D$-- meson which was calculated by formula
(\ref{r2n}).  The fitting gives the following values: in the
model corresponding to Eq. (\ref{HO-wf}) we obtain
$b=0{.}372\,$ GeV,  in the model
(\ref{PL-wf}) we find $b=0{.}526\,$ GeV for $n=$2  and
$b=0{.}731\,$ GeV for
$n=$3 .
With the
presented wave functions
the following values
for slope parameter in Eq.  (\ref{ourrho})
were obtained: $\rho^2=1{.}32$  in the model (\ref{HO-wf}),
$\rho^2=1{.}13$ in the model (\ref{PL-wf})
for $n=$2 and $\rho^2=1{.}20$ for  $n=$3 , and
$\rho^2=1{.}20$ in the model
(\ref{Tez91-wf}) .

The result in model of harmonic-oscillator satisfies
inequalities (\ref{rhovol}) within the accuracy of our calculations.
Let us remind in this connection that the restrictions (\ref{rhoint})
were obtained for the model with potential increasing
slowlier than linearly.
It is possible that the restrictions (\ref{rhoint})
are valid for wider class of models.
The calculations with exact formulae (\ref{ourrho}),
(\ref{rhos}) for models (\ref{Tez91-wf}), (\ref{PL-wf}) satisfy
the restriction (\ref{rhovol}).

Let us compare our results (\ref{rhovol}), (\ref{soft}) with recent
experimental values of $\rho^2$. Result of CLEO Collaboration \cite{CLEO},
\cite{Bat00}
($\rho^2$ = 1.67$\pm$0.11(stat)$\pm$0.22(syst)) satisfies our "soft"
result (\ref{soft}) and is slightly outside the region
(\ref{rhovol}).
The average LEP result \cite{LEP},
\cite{Bat00} ($\rho^2$ = 1.13$\pm$0.08(stat)$\pm$0.16(syst))
satisfies the "hard" restriction (\ref{rhovol}).

Let us consider results for $\rho^2$ in other
approaches.
The two restrictions on slope parameter $\rho^2$ were obtained
in QCD sum rules:
$\rho^2 \geq 0{.}25$ \cite{Bjo90} , $\rho^2 \leq 1{.}7$ \cite{RafTa94}.
Quasipotential approach gives $\rho^2\ge$ 0{.}81 \cite{EbF00}.
Another restriction was obtained in the framework of RHD
\cite{YaoOl9596} $\rho^2 \geq 3/4$.
As we see, our restriction (\ref{rhovol}) does not contradict
these results but reinforces them.
However, our result, as well as that of work \cite{YaoOl9596} and experiments
\cite{CLEO}, \cite{LEP}, is in poor correlation
with the result of Ref. \cite{Vol92} obtained in the framework
of QCD sum rules $\rho^2 \leq  0{.}75\pm 0.15$.

Now let us consider the calculations in the framework
of instant form of RHD.
In work \cite{MoYaOl9697}, the calculations are performed for different
interaction models of
quarks in meson and results are within the range from 0.97 to
1.28, that is very close to
(\ref{rhovol}).
The result of quasipotential approach
\cite{FaG95} coincides with that of ours, $\rho^2 = 1{.}02$.

The results of different calculations of $\rho^2$ in
the framework of QCD sum rules differ  considerably:  $
1{.}00\pm 0{.}02\, ~\cite{Nar94}\,~, 0{.}84\pm 0{.}02\,
~\cite{BaB93}\,~, 0{.}66\pm 0{.}05\, ~\cite{Neu9293}\,~,
0{.}70\pm 0{.}25\, ~\cite{BlS93}\,~.
$
First and fourth of these values are near our range (\ref{rhovol}).

The lattice calculations \cite{Bow95} are in good correlation with our result,
but we should point the large errors of these calculations:
$\rho^2 = 0.9^{+0.2+0.4}_{-0.3-0.2}$.

Now let us discuss briefly the calculation of IWF in our approach.
In fig.1 the results of calculation for IWF with
the described model wave functions are shown.
The calculation was carried out with the parameters of this work at $m_q$=0.25
GeV.
As can be seen from the fig.1, one of the most interesting
features of our calculations is the weak dependence of IWF on
the choice of model wave functions at $w\>\le$2.
Hence, the properties of IWF are mainly determined by the
relativistic kinematics of light quark or so-called "light" degrees
of freedom.

So, in the present work, the Isgur -- Wise
function is calculated in the framework of the instant form of
relativistic Hamiltonian  dynamics.  A strong model--independent
restriction is obtained for the slope parameter of Isgur -- Wise
function.  This restriction reinforces the restrictions that
were given in other approaches.  We have also shown that Isgur
-- Wise function depends weakly on the choice of the model wave
functions at $w\>\le$2.

We are grateful to A.V.Gorokhov for numerous useful discussions.
This work was supported in part by the program "Russian Universities --
Basic Researches" (grant N 02.01.28).

\pagebreak


\pagebreak

{\bf Figure captions}\\[1cm]

Fig. 1. Results of the calculations of Isgur--Wise function with
the different model wave functions. Dotted line: the power law
wave function of Eq. (\ref{PL-wf}) with
$n=2$; dashed line:  the same model with $n=3$; dot--dashed line:
the wave function in the model with linear confinement and
Coulomb behavior at small distances of Eq.  (\ref{Tez91-wf});
solid line: harmonic--oscillator wave function of Eq. (\ref{HO-wf}).
Wave function parameters were determined by fitting of mean
square radius of $D$-- meson calculated with formula (\ref{r2n}).

{~}


\begin{figure}[htbp]\vspace*{-3.0cm}
\epsfxsize=0.9\textwidth
\centerline{\psfig{figure=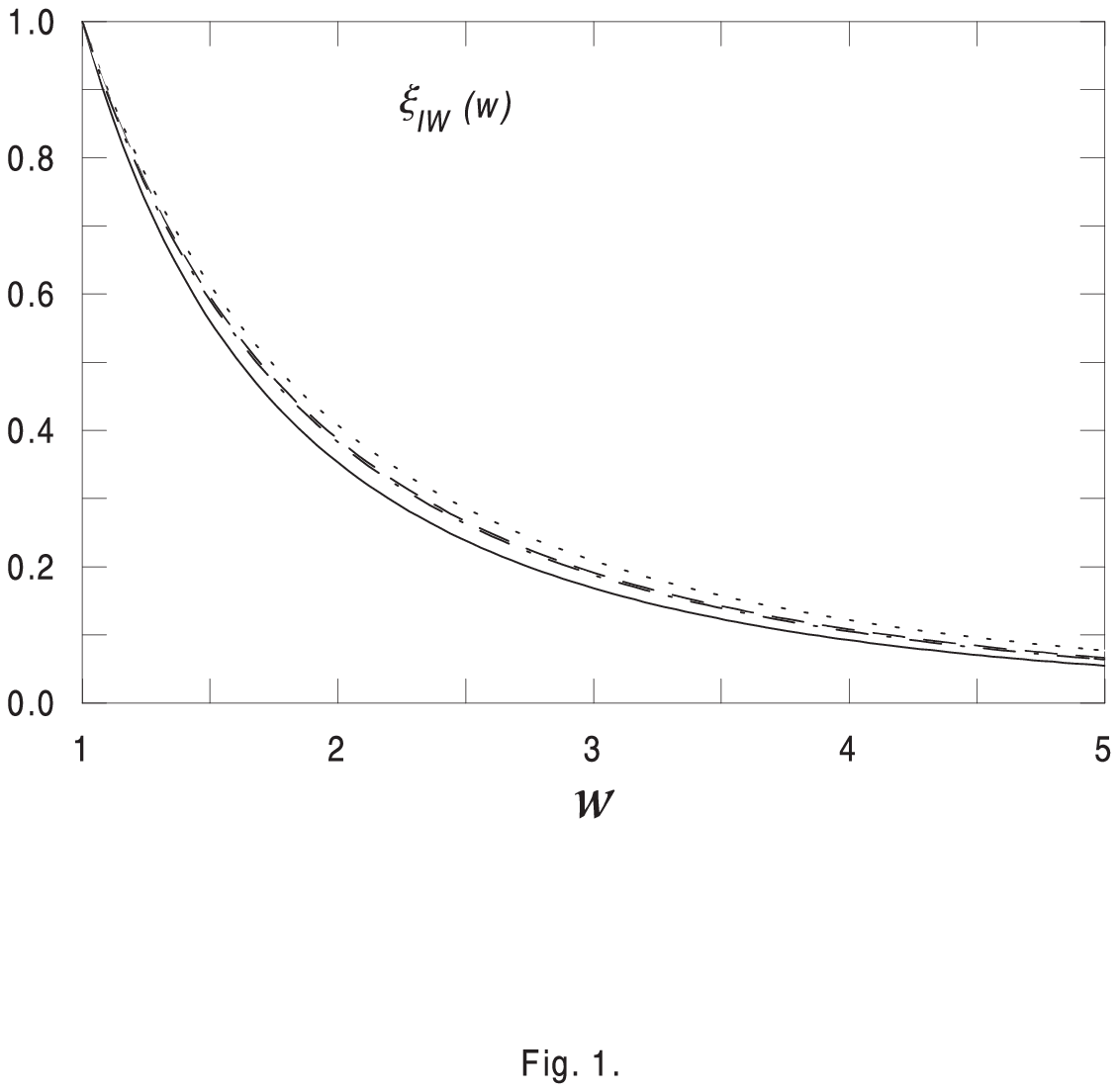,height=30cm,width=20cm}}
\vspace{0.3cm}
\begin{center}
{\bf е}
\end{center}
\vspace*{-2.5cm}
\end{figure}

\end{document}